\documentclass[twocolumn]{aastex6}

\fullcollaborationName{The Friends of AASTeX Collaboration}

\begin{document}


\title{Radial velocities and metallicities of red giant stars in the old open cluster NGC 7762}


\author{Giovanni Carraro\altaffilmark{1,2,3}}
\affil{Dipartimento di Fisica e Astronomia {\it Galileo Galilei}, \\
Vicolo Osservatorio 3,\\
I-35122, Padova, Italy}

\author{Eugene A. Semenko}
\affil{Special Astrophysical Observatory, Russian Academy of Sciences,\\
 Nizhny Arkhyz 369167, Russia}

\and

\author{Sandro Villanova}
\affil{Departamento de Astronom\'ia, Universidad de Concepci\'on,\\
 Casilla 160-C, Concepci\'on, Chile}

\altaffiltext{1}{ESO, Alonso de Cordoba 3107, Santiago de Chile, Chile}
\altaffiltext{2}{giovanni.carraro@unipd.it}
\altaffiltext{3}{Visiting Astronomer, Special Astrophysical Observatory}

\begin{abstract}
We present and discuss radial velocity and the very first metallicity measurements for nine evolved stars in the poorly known old open cluster NGC 7762.  We isolated eight radial velocity cluster members and one interloper. Radial velocities are in good agreement with previous studies. NGC 7762 turns out to be of solar metallicity within the uncertainties ([Fe/H]=0.04$\pm$0.12). For this metallicity, the cluster age is 2.5$\pm$0.2 Gyr, and falls in a age range where only a few old open clusters are known. With  respect to previous studies, we find a larger distance, implying the cluster to be located at 900$^{+70}_{-50}$ pc from the Sun. 
For most of the elements we measure solar-scaled abundance ratios. We searched the literature for open clusters of similar age in the solar vicinity and found that NGC 7762 can be considered a twin of Ruprecht 147, a similar age cluster located at only 300 pc from the Sun. In fact, beside age, 
also metallicity and  abundance ratios  are very close to Ruprecht 147 values within the observational uncertainties. 
\end{abstract}


\section{Introduction} \label{sec:intro}
The last decade witnessed a significant accumulation of good quality, high-resolution, spectroscopic metallicity measurements for old open clusters in the Galactic disk. 
Metallicity, together with abundance ratios for a large sample of elements, of old open clusters at different age and location in the disk are the natural
constraints for chemical evolution models that aim at understanding how the disk of the Milky Way formed and evolved.
Essentially, this is because old open clusters span a wide range of ages, and are almost ubiquitous in the disk, thus allowing to build up an observational age metallicity relation, the actual radial abundance gradients, and their evolution in time (Magrini et al. 2009).
Additionally, open clusters offer the obvious advantage that the abundance of elements can be derived in statistical fashion, since no evidence has been found so far that they deviate from the single stellar population concept.\\

\noindent
In this work, we present a spectroscopic investigation of the old open cluster NGC 7762, for which no abundance analysis has been performed to date. \\
The cluster was studied for the first time by Chincarini (1966), who in a short note provided photographic UBV photometry of 18 stars, and preliminarily suggested a distance of 1 kpc and a reddening E(B-V)= 1.02. The same author recommended
to obtain more data to extend the cluster main sequence (MS) for fainter magnitudes. Zakharova (1972) obtained an independent photographic data-set, of similar photometry depth, and derived a distance of 890 pc and a reddening E(B-V)=0.89. She provided also an estimate of the mass ($\sim 400 M_{\odot}$, and argue that the cluster is similar to the more famous NGC~752, NGC~1245, and NGC~7789, which implies an age above 1 Gyr. More recently,  Patat \& Carraro (1995) acquired the very first CCD photometry, extending the MS down to V$\sim$ 19.0. They noticed, however, that the MS  disappears at V $\sim$ 16.5, and interpreted this as an evidence of low-mass star evaporation. In the same study they derive an age of 1.8 Gyr, a distance of 800 pc, and a reddening  E(B-V) $\sim$0.85.
The most recent investigation is by Maciejewski et al. (2008), where the short distance derived previously is confirmed, but the derived age is larger (2.4 Gyr), and the reddening significantly smaller (E(B-V) =0.59).
Clearly, a significant improvement of the cluster basic parameters can be obtained if metallicity is known.\\
We observed nine stars in the cluster red clump and derived their radial velocity and metallicity. The selected stars are shown in the color magnitude diagrams (CMD) in Fig.~1,  and constructed using Maciejewski et al. (2008) photometric data.
They are highlighted in red, and their properties are summarised in Table~1.\\

\noindent
The paper is organised as follows: in Sect. 2 we present the observational material and the basic reduction strategy. 
Section 3 is devoted to radial velocities, while in Sect. 4 we discuss the derivation of the stars' atmospheric parameters.  The abundance analysis is presented in Section~5. Section~6 is then dedicated to an update of the cluster fundamental parameters. In Section~7 we discuss our results and compare our abundance ratio determinations with the literature. Finally, Section~8 summarises our findings.

\begin{table*}
\tabcolsep 0.3truecm
\caption{Basic information of the stars for which we obtained spectroscopic data. IDs are taken from Maciejewski et al. (2008).The $\frac{S}{N}$  in the last column has been measured using a few \AA~ continuum region at 6300\AA.}
\begin{tabular}{lccccccccl}
\hline
ID &	RA(2000.0)  &    Dec(2000.0)	&  V	& B-V  &    J  &    H  &    K  &   exp (sec) &    $\frac{S}{N}$ \\
\hline\hline
361	 & 23:48:54.4 &  68:03:52.1  & 11.956 & 1.704  &  8.813  & 8.137  & 7.980  &   2400 &  80\\ 
451	 & 23:49:06.1 &	 67:59:08.6  & 12.559 & 1.712  &  9.291  & 8.636  & 8.419  &   2700 &  65\\
459	 & 23:49:07.0 &	 67:55:24.2  & 12.366 & 1.690  &  9.108  & 8.436  & 8.286  &   2700 &  70\\
535	 & 23:49:15.7 &	 68:05:32.1  & 12.569 & 1.693  &  9.257  & 8.581  & 8.372  &   2700 &  65\\
820	 & 23:49:48.4 &	 68:01:35.1  & 12.564 & 1.758  &  9.172  & 8.496  & 8.272  &   2700 &  70\\
831	 & 23:49:49.3 &	 68:01:07.3  & 12.876 & 1.796  &  9.232  & 8.543  & 8.261  &   3600 &  70\\
1195	 & 23:50:31.4 &	 68:01:41.5  & 11.730 & 1.751  &  8.431  & 7.781  & 7.518  &   3600 & 110\\
1387	 & 23:50:59.3 &	 68:00:36.6  & 12.802 & 1.786  &  9.350  & 8.682  & 8.424  &   3600 &  70\\
1455	 & 23:51:12.1 &	 67:56:29.7  & 11.479 & 2.046  &  7.594  & 6.719  & 6.469   &  1800 &  90 \\ 
\hline\hline 
\end{tabular}
\end{table*}

\section{Observation and Data Reduction} \label{sec:observation}
We observed NGC 7762 stars with the Main Stellar Spectrograph (MSS)\footnote{www.sao.ru/hq/lizm/mss/en/index.html} of the 6 meter telescope BTA at the Special Astrophysical Observatory in Nizhny Arkhyz, Russia. This instrument is installed in the Nasmyth 2 focus of the telescope. It is essentially a long-slit spectrograph, which in its standard mode  is equipped with a differential circular polarisation analyser, and  combined with a double image slicer designed by Chountonov (2004) and a rotating l/4 phase plate, to study mainly stellar magnetic fields.
For the purpose of our observations, the instrument was set-up in its basic  long slit mode. In this set-up, the MSS instrument allows to cover the wavelength range 5940$-$6690\AA~ with a resolution $\lambda / \Delta \lambda$ of $\sim$ 13000 .\\
Observations were carried out on the night of August 12, 2016. The night was clear and stable, with typical seeing of $\sim$ 1 arcsec.
Beside the spectra of scientific targets (one exposure per target), we collected bias and dome flat frames (two sets of 10 images obtained at the beginning and at the end the night) and a set of ThAr spectra, for wavelength calibration purposes.  Our intention was to observe all the stars in the magnitude range $11 \leq V \leq 13$, and redder than (B-V) $\sim$ 1.6, but could obtained spectra only for nine stars in the allocated time.\\

\noindent
Spectra were reduced in a standard  way, which consists of the following steps: 1) construction of the master bias and subtraction of it from the scientific and calibration frames, 2) correction of data for stray light, 3) searching for the location of individual slices in 2D images, 4) extraction of 1D spectra, 5) continuum normalisation of the spectra, 6) correction of wavelengths for the Earth's motion. All these procedures were carried out within ESO-MIDAS package and its Zeeman extension. Spectra were normalised by the continuum level using the task  {\it continuum} within the IRAF package\footnote{IRAF is distributed by the National
Optical Astronomy Observatory, which is operated by the Association of
Universities for Research in Astronomy, Inc., under cooperative agreement
with the National Science Foundation}. Wavelength solution was defined using  the arc spectrum obtained closest in time to the scientific exposures.\\
An excerpt of the nine stars' spectra is shown in Fig.~2.

\begin{figure}
\includegraphics[width=\columnwidth]{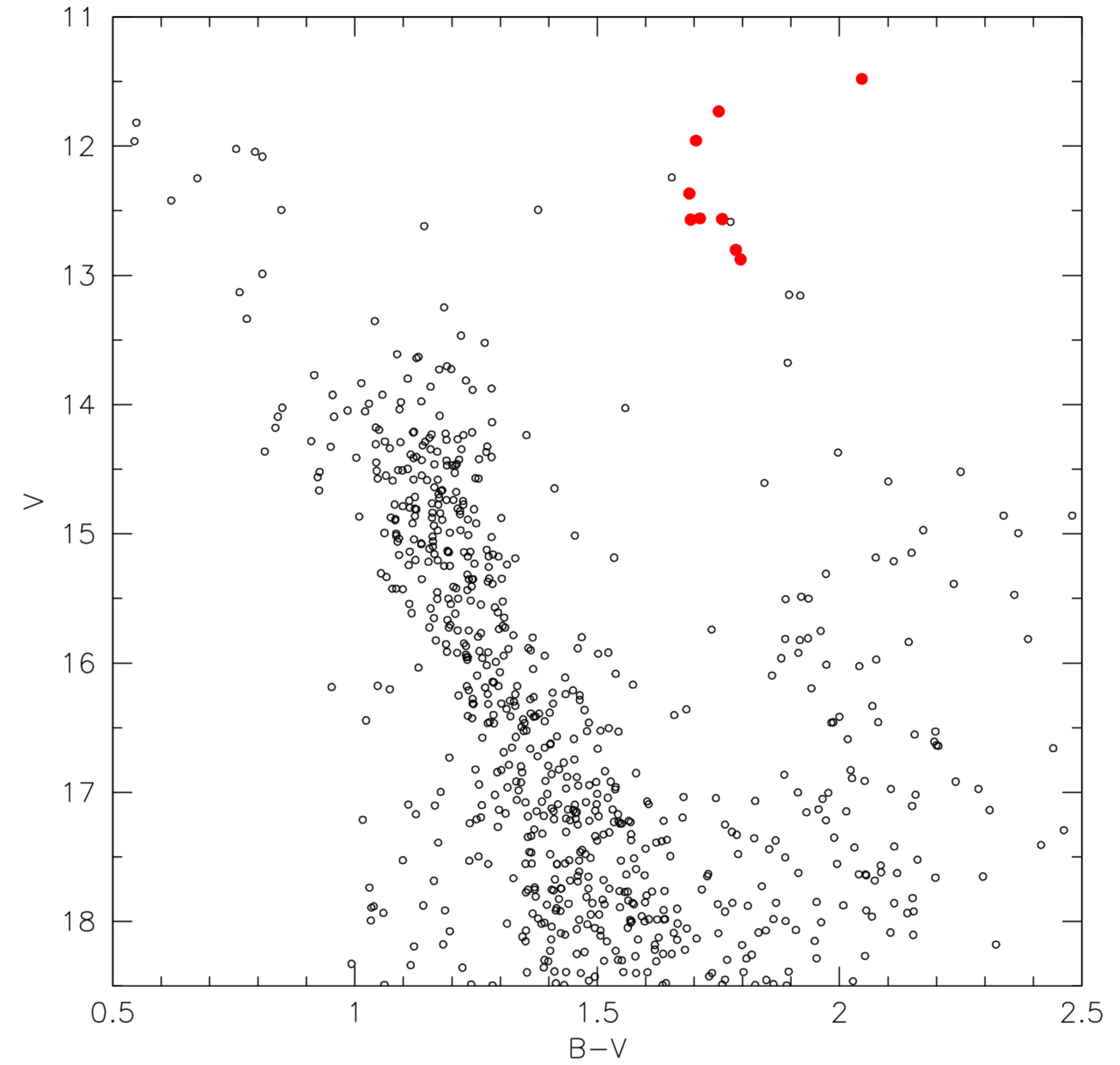}
\caption{Color magnitude diagram of NGC 7762. Red symbols indicate the stars for which we obtained spectroscopic data. Photometry is taken from Maciejewski et al. (2008).}
\end{figure}

\section{Radial velocities}
Radial velocities were measured by the {\it fxcor} package in IRAF, using a synthetic spectrum as a
template calculated for a typical giant star with solar metallicity
(T$_{eff}$=4800 K, log(g)=2.50, v$_{t}$=1.50 km/s). All stars except for $\#361$ turned
out to have the same velocity within few km/s,  and were considered members. 
The mean radial velocity for the cluster is -46.5$\pm$0.8 km/s. The typical
error on radial velocity given by  {\it fxcor} is 1.0 km/s.\\
We compare in Table~2 our results with Casamiquela et al. (2016), who, in the context of the OCCASO (Open Cluster Chemical Abundances from Spanish Observatories) survey report radial velocity measurements for 6 stars in NGC 7762.

\begin{table}
\tabcolsep 0.2truecm
\caption{Radial velocity measurement for the nine program stars in NGC 7762. Literature values are from Casamiquela et al. (2016).}
\begin{tabular}{lcccc}
\hline
ID &	RA(2000.0)  &    Dec(2000.0) & $V_R$  & $V_{R, OCCASO}$ \\
\hline
 & & & [km/s] & [km/s] \\
361	 & 23:48:54.4 &  68:03:52.1  &   36.4   &   \\ 
451	 & 23:49:06.1 &	 67:59:08.6  &  -45.6   &  -45.9$\pm$0.7 \\
459	 & 23:49:07.0 &	 67:55:24.2  &  -48.8   &   \\
535	 & 23:49:15.7 &	 68:05:32.1  &  -47.9   &  -45.5$\pm$0.6  \\
820	 & 23:49:48.4 &	 68:01:35.1  &  -46.2   &  -45.7$\pm$0.7 \\
831	 & 23:49:49.3 &	 68:01:07.3  &  -44.4   &  -47.0$\pm$0.8 \\
1195	 & 23:50:31.4 &	 68:01:41.5  &  -50.2   &   \\
1387	 & 23:50:59.3 &	 68:00:36.6  &  -45.4   &  -45.6$\pm$0.8 \\
1455	 & 23:51:12.1 &	 67:56:29.7  &  -43.8   &   \\ 
\hline\hline 
\end{tabular}
\end{table}

\noindent
We have five stars in common with the OCCASO survey, and the comparison is extremely good (see Table~2).

\begin{table}
\tabcolsep 0.4truecm
\caption{Atmospheric parameters for member stars.}
\begin{tabular}{lccc}
\hline\hline
ID   & $T_{eff}$ &  logg &  $v_t$\\
\hline
 & [$^o$K] & & [km/s] \\
\hline
451   & 4800  &  2.655   & 1.323\\
459   & 4845  &  2.601   & 1.424\\
535   & 4839  &  2.679   & 1.347\\
820   & 4713  &  2.609   & 1.260\\
831   & 4640  &  2.692   & 1.098\\
1195 & 4726  &  2.283   & 1.569\\
1387 & 4659  &  2.674   & 1.137\\
1455 & 4214  &  1.833   & 1.365\\
\hline\hline 
\end{tabular}
\end{table}

\section{Atmospheric parameters}

Atmospheric parameters were obtained as follows. 
First, T$_{\rm eff}$ was derived from the B-V color using the relations by
Alonso et al. (1999) and Ramirez \& Melendez (2005).
Surface gravities (log(g)) were obtained from the canonical equation:

\begin{center}
{\rm log($\frac{g}{g_{\odot}}$) = log($\frac{M}{M_{\odot}}$) + 4$\cdot$
log($\frac{T_{\rm eff}}{T_{\odot}}$) - log($\frac{L}{L_{\odot}}$)}
\end{center}

\noindent
The bolometric correction was derived by adopting the relations 
from Alonso et al. (1999) and Flower (1996).
We adopted initial values for reddening E(B-V) and distance modulus ${(m-M)_{\rm V}}$ from WEBDA and inferred
stellar masses from isochrones (Bressan et al. 2012). We found  E(B-V)=0.71 and ${(m-M)_{\rm  V}}$=11.56, and assumed a 
typical mass of 1.50 M$_{\odot}$. 
Micro-turbulent velocity (v$_{\rm t}$) was obtained from the
relation of Gratton et al. (1996)  that considers both temperature and gravity as
parameters:\\

\begin{center}
{\rm v$_{\rm t}$=0.00119$\cdot$T$_{\rm eff}$-0.90$\cdot$log(g)-2}
\end{center}

\noindent
The input metallicity needed to obtain T$_{\rm eff}$ from colours and for the
isochrone fitting was assumed to be solar, which was later confirmed by the spectroscopic analysis (see below). 
The LTE program MOOG (Sneden 1973) was used
for the abundance analysis coupled with atmosphere models by Kurucz (1970).
We adopted the same line-list adopted in all our previous papers (Carraro et al. 2014a,b).

\section{Elemental abundances}
SiI, CaI, TiI, FeI, FeII and NiI abundances were estimated
using the equivalent width (EQW) method. EQWs were obtained fitting a gaussian
to the spectral features.
NaI and BaII abundances were obtained using the spectro-synthesis method. For this
purpose 5 synthetic spectra were generated for each line with 0.25 dex
abundance step and compared with the observed spectrum. The 
line-list and the methodology we used are the same used in previous papers (e.g. Villanova et al. 2013),
so we refer to those articles for a detailed discussion of this point. 
Here we would like to stress that to measure Ba abundance hyperfine splitting was taken into account, 
as in our previous studies (e.g. Carraro et al. 2014a,b). The derived elemental abundances are reported in Table~4.\\

\begin{figure}
\includegraphics[width=\columnwidth]{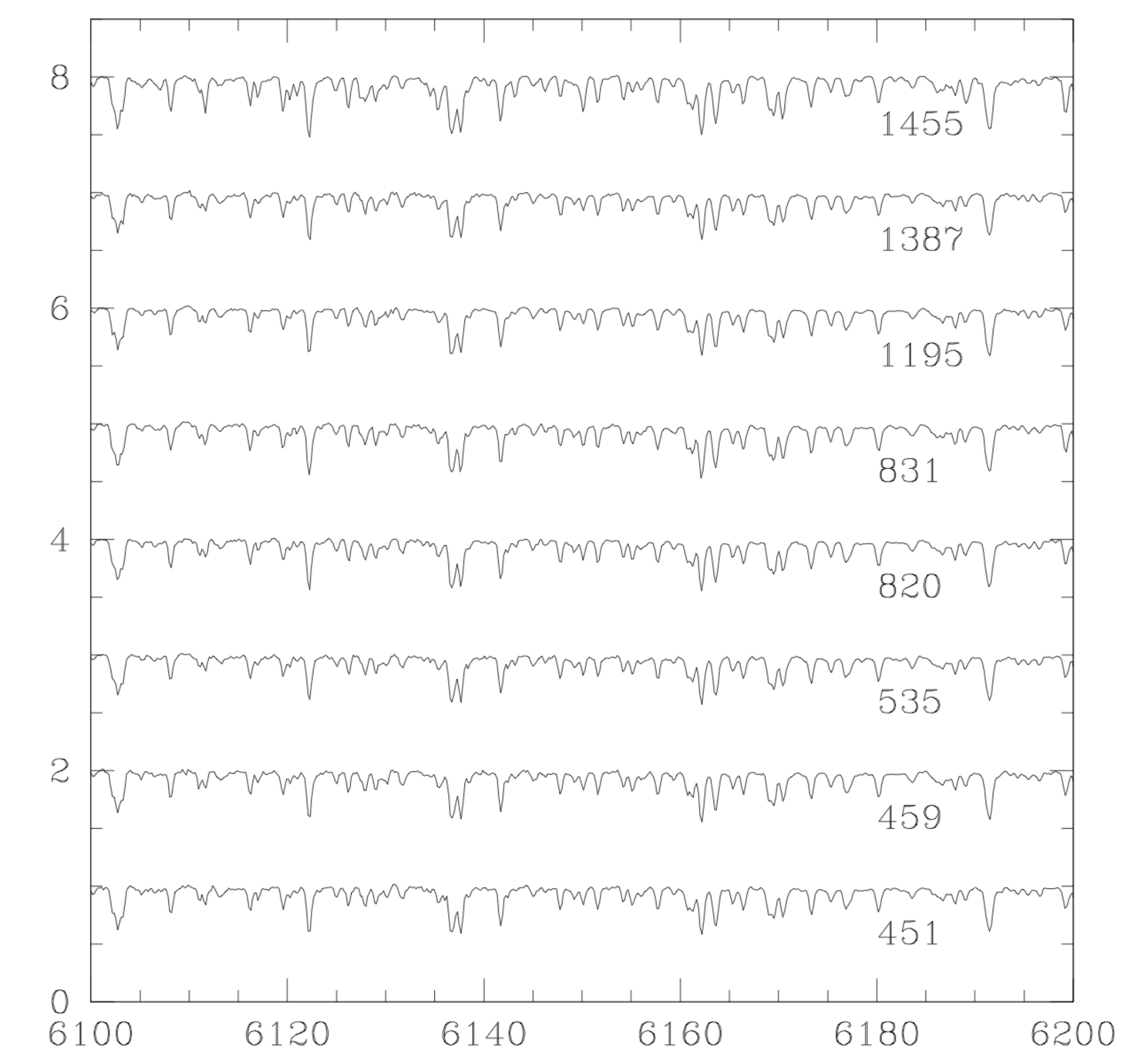}
\caption{An excerpt of the nine individual spectra in the indicated wavelength range. Spectra are arbitrarily shifted in the vertical axis, for illustration purposes.}
\end{figure}

\noindent
Using FeI and FeII we were able to cross-check the atmospheric parameters
obtained from the photometry in the following way. First of all for each star
we calculated the mean Fe abundance from FeI lines. Then we calculated for
each FeI and FeII line the difference $\Delta(Fe)$ with respect the mean value.
Finally we reported these differences in Fig.~3.
for all the stars. In the upper panel we plot $\Delta(Fe)$ as a function of
the excitation potential E.P. and in the lower panel as a function of the
reduced equivalent width\footnote{The reduced equivalent width (R.E.W.) is defined as $log_{10}(\frac{Equivalent  Width}{\lambda})$ .}.
We used black points for FeI lines and red 
points for FeII lines. In both panels FeI lines were fitted with a straight blue
line. The slope of each line is given in the upper part of each panel. Both
slopes are compatible with a flat trend within 1$\sigma$. This means that 
temperature and micro-turbulence scales we are using are reliable.
In the same way we report in the upper panel the difference between the mean
FeI and FeII abundances. Also in this case the difference is
compatible with zero within 1$\sigma$ and we conclude that the gravity scale
we are using is reliable.

\noindent
In order to estimate the uncertainties associated with these estimates, we considered star $\#1387$, and repeated the
same process as above at varying  $T_{eff}$ by 50$^oK$, log(g) by 0.20, $v_T$ by 0.1 km/s , and [Fe/H] by 0.1 dex. 
Table~5 summarises our results. The dependence on the limited spectra $\frac{S}{N}$ is also show. The column indicated with 
{\it Tot} is the quadratic sum of the various contributions,  while the term {\it Obs} indicates the dispersion of the ratio measurements among the 8 
member stars.

\begin{table}
\tabcolsep 0.05truecm
\caption{Abundance analysis for member stars.}
\begin{tabular}{lcccccccc}
\hline\hline
ID  &   [Fe/H] & [Si/Fe] & [Ca/Fe] &[ TiI/Fe] & [Ni/Fe] & [Na/Fe] & [Ba/Fe]\\
451   &  0.08 &  0.08 &  0.18  & 0.00  &  0.17 &-0.26 & 0.32\\
459   &  0.09 &  0.25 &  0.26  & 9.99  &  0.17 &-0.14 & 0.38\\
535   & -0.02 &  9.99 &  0.06  & 0.10  &  0.11 &-0.19 & 0.37\\
820   &  0.09 &  0.34 &  0.05  &-0.02  &  0.15 &-0.29 & 0.47\\
831   &  0.18 &  0.23 &  0.04  & 9.99  &  0.29 &-0.28 & 0.45\\
1195 & -0.15 &  0.29 &  0.11  & 0.02  &  9.99 & 0.03 & 0.43\\
1387 & -0.12 &  0.34 &-0.10  &-0.06  &  0.20 &-0.06 & 0.60\\
1455 &  0.15 &  9.99 &  9.99 & 0.26   &  0.09 &-0.36 & 0.44\\
\hline\hline 
\end{tabular}
\end{table}

\begin{figure}
\includegraphics[width=\columnwidth]{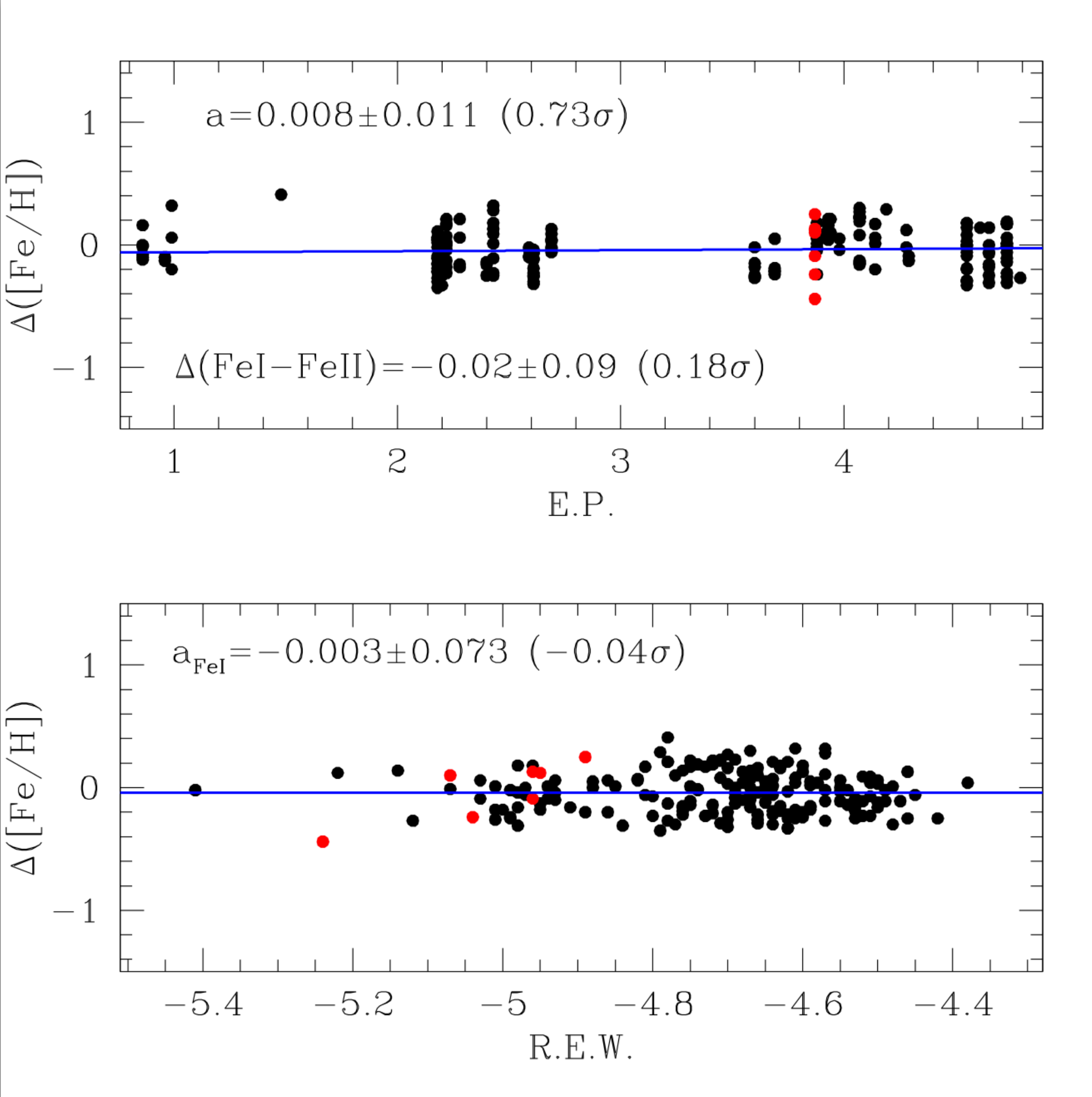}
\caption{Difference of [Fe/H] measurements from FeI and FeII with respect to the mean Fe abundance a a function of the excitation potential (E.P.)
and the reduced equivalent width (R.E.W.). See text for details.}
\end{figure}

\begin{table}
\tabcolsep 0.15truecm
\caption{Error budget. Each column reports the error dependence on the various atmospheric parameters. }
\begin{tabular}{lccccccc}
\hline\hline
Ratio &   $T_{eff}$ & log(g)  &$v_t$  & [Fe/H]  & $\frac{S}{N}$ & $Tot$ &  $Obs$\\
\hline
$[Si/Fe]$  &  0.04  & 0.03 & 0.03 & 0.03 & 0.05 & 0.08 &  0.10 \\
$[Ca/Fe]$ &  0.05  & 0.02 & 0.05 & 0.02 & 0.07 & 0.10 &  0.11 \\
$[Ti/Fe]$   &  0.05  & 0.01 & 0.02 & 0.01 & 0.08 & 0.10 &  0.12 \\
$[Fe/H]$   &  0.06  & 0.02 & 0.07 & 0.02 & 0.03 & 0.10  & 0.12 \\
$[Ni/Fe]$  &  0.02  & 0.02 & 0.01 & 0.01 & 0.07 & 0.08  & 0.07 \\
$[Na/Fe]$ &  0.02  & 0.03 & 0.05 & 0.01 & 0.08 & 0.10  & 0.13 \\
$[Ba/Fe]$ &  0.02  & 0.06 & 0.05 & 0.02 & 0.03 & 0.09  & 0.08 \\
\hline\hline 
\end{tabular}
\end{table}

\section{Cluster basic parameters revisited}
The abundance analysis carried out in previous sections yields a mean metal abundance [Fe/H]=0.04$\pm$0.12, compatible with the solar value. From the individual values (see Table~4) one can notice some spread, which, however, is totally compatible with the observational errors. Using this new, spectroscopic, metal abundance estimate, we can now derive NGC 7762 fundamental parameters, and for this purpose we employ the Padova suite of stellar models and isochrones (Bressan et al. 2012).  Our results are summarised in Fig.~4. We favor an age of 2.5$\pm$0.2 Gyr. The corresponding isochrones nicely fit both the cluster MS and the red giant clump. This solution implies a reddening E(B-V) =0.62$\pm$0.03 and an apparent distance modulus (m-M)$_V$ = 11.7$\pm$0.1. Uncertainties are estimated by visual inspection,  shifting the isochrone in the horizontal and vertical direction until the fit becomes unacceptable.\\

\noindent
While our reddening value is compatible with Maciejewski et al. (2008) estimate, we find some disagreement ($\approx10\%$ larger) for the distance modulus estimate,
despite the fact we are using their photometry and roughly  the same age. However, if one considers carefully their isochrone solution (their Fig.~3), it is immediately clear that the fit is poor both in optical and in infrared. In fact, the optical fit completely misses the clump, while the infrared fit matches the clump position, but not the cluster TO. \\
The distribution of the observed stars in the CMD deserves in fact more attention.  We have adopted as position of the red giant clump the location of the three stars $\# 451$, $\#535$, and $\#820$, that have the same magnitude level. According to this solution, we would argue that the two fainter and slightly redder stars $\#831$ and $\#1387$ are red giant branch (RGB) stars, while the remaining stars are most probably asymptotic giant branch (AGB) stars, which are expected to be present in the cluster of this age. The isochrone  mismatch 
in color is not particularly worrisome, given the various uncertainties affecting models in these evolutionary phases: mixing length parameter calibration, transformation from theoretical to observational plane, atmosphere models, and so forth.\\

\noindent
Our new distance estimate then places NGC 7762 at 900$^{+70}_{-50}$ pc from the Sun.\\

\noindent
This updated set of parameters (reddening and apparent distance modulus) slightly differ from the one we adopted from WEBDA as input to estimate  stars' atmospheric parameters. 
We investigated this issue in details and found that there is no implication for $log(g)$, since a difference in apparent distance modulus of a few hundredths magnitudes only (within the errors)
would produce a difference in $log(g)$ of 0.07. Reddening is on the other hand critical for the derivation of $log (T_{eff})$. In our case, if we take errors into consideration
the difference in E(B-V) is 0.06 mag (1 $\sigma$ difference), which causes a difference in $log (T_{eff})$ of 80 degrees. Stars would then be slightly cooler, and their metallicity slightly
lower (by 0.05 dex ), which on the other hand is largely compatible with our estimate.

\begin{figure}
\includegraphics[width=\columnwidth]{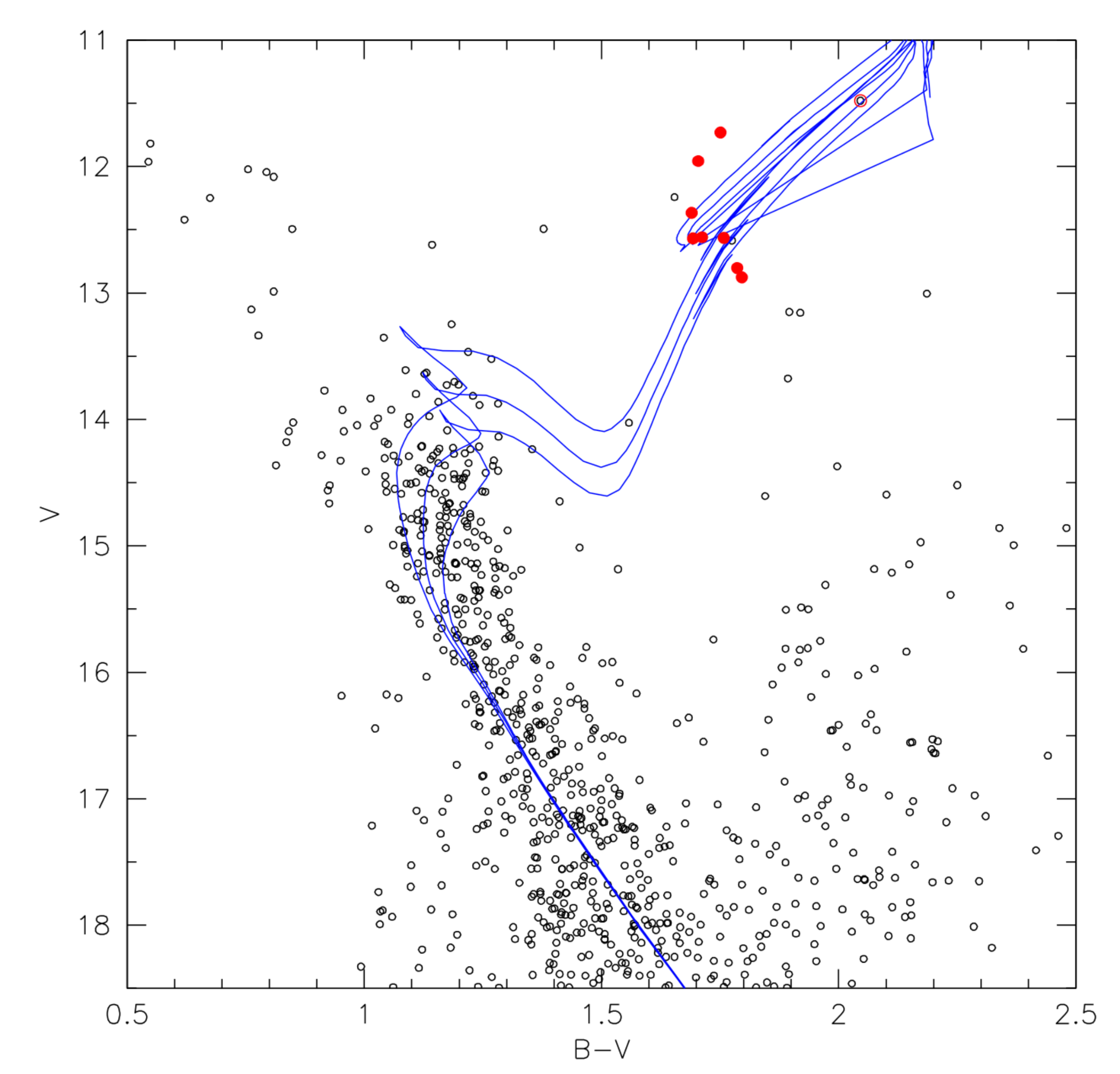}
\caption{NGC 7762 basic parameters' solution. Isochrones in blue are shown for ages of 2.0, 2.5 and 3.0 Gyr, and super-imposed for E(B-V) = 0.62, and (m-M) = 11.7. Red filled symbols indicate the member stars for which spectroscopy has been obtained in this study. The empty symbol refers to the non-member.}
\end{figure}

\section{Discussion}
Beside metal abundance, we derived also a few abundance ratios, as listed in Table~4. Their uncertainties are listed in Table~5. We are going to compare our findings with the population of Galactic old open clusters for which such ratios are available, to put NGC 7762 into context.\\

\subsection{Metallicity and $\alpha$ elements}
With an age of 2.5 Gyr and a metallicity of [Fe/H]=0.04$\pm$012, NGC 7762 looks as a typical old open cluster in the solar neighbourhood.  There are not many old open clusters  closer than $\sim$ 1 kpc to the Sun, to compare our findings with. If we generate such list from WEBDA\footnote{http://www.univie.ac.at/webda/clusterselall.html}   we find that only one cluster sharing the same age as NGC~7762: Ruprecht 147 (Curtis et al. 2013). According to these authors, Ruprecht 147 is also a 2.5$\pm$0.25 Gyr old open cluster . Both clusters, by the way, are quite spread over the sky,
being this a combined effect of their vicinity to the Sun and the advanced stage of dynamical evolution. Interestingly, Ruprecht 147 has a metallicity of 
[Fe/H]=0.07$\pm$0.03, again virtually identical to NGC 7762. \\

\begin{figure*}
\plottwo{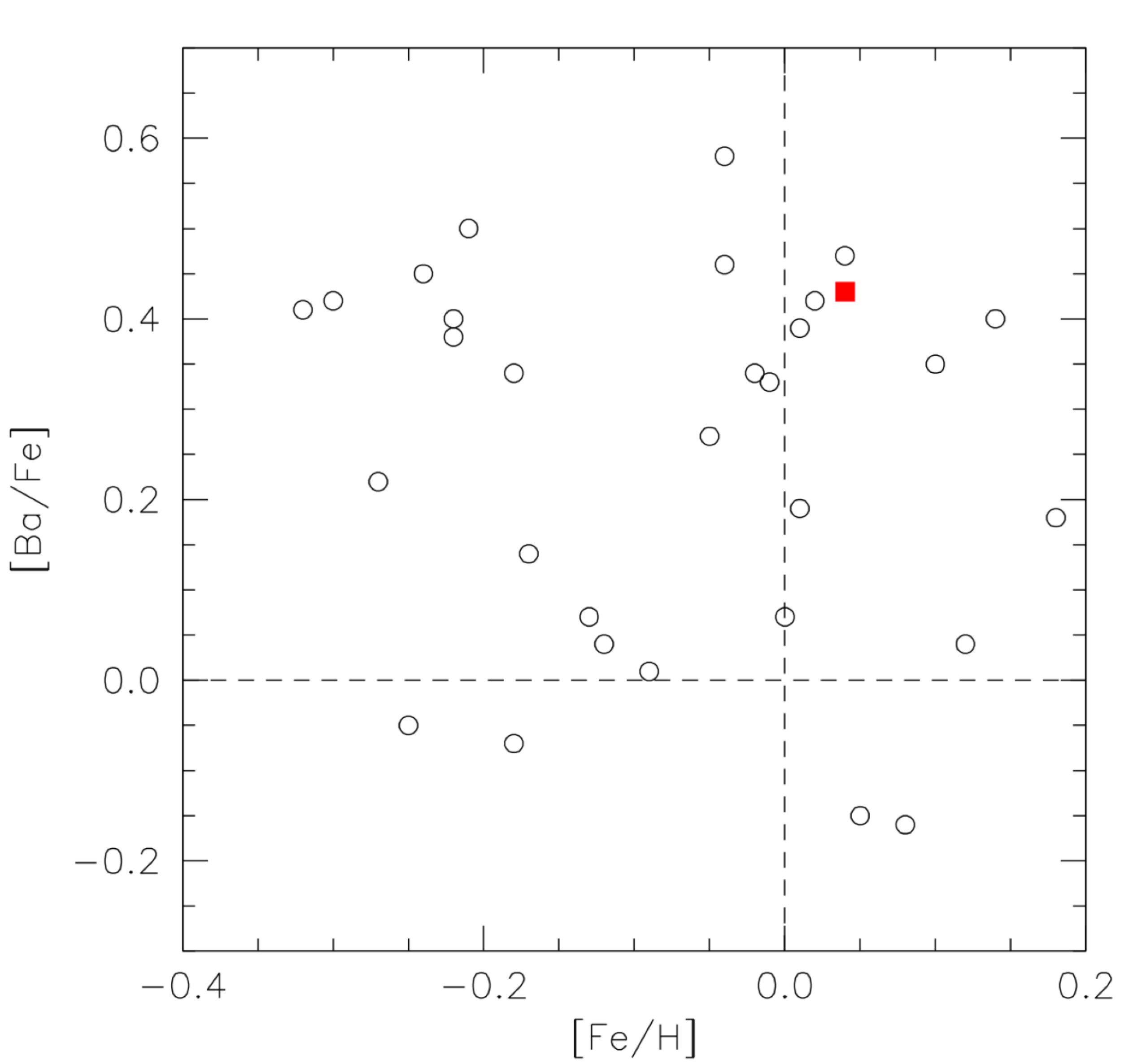}{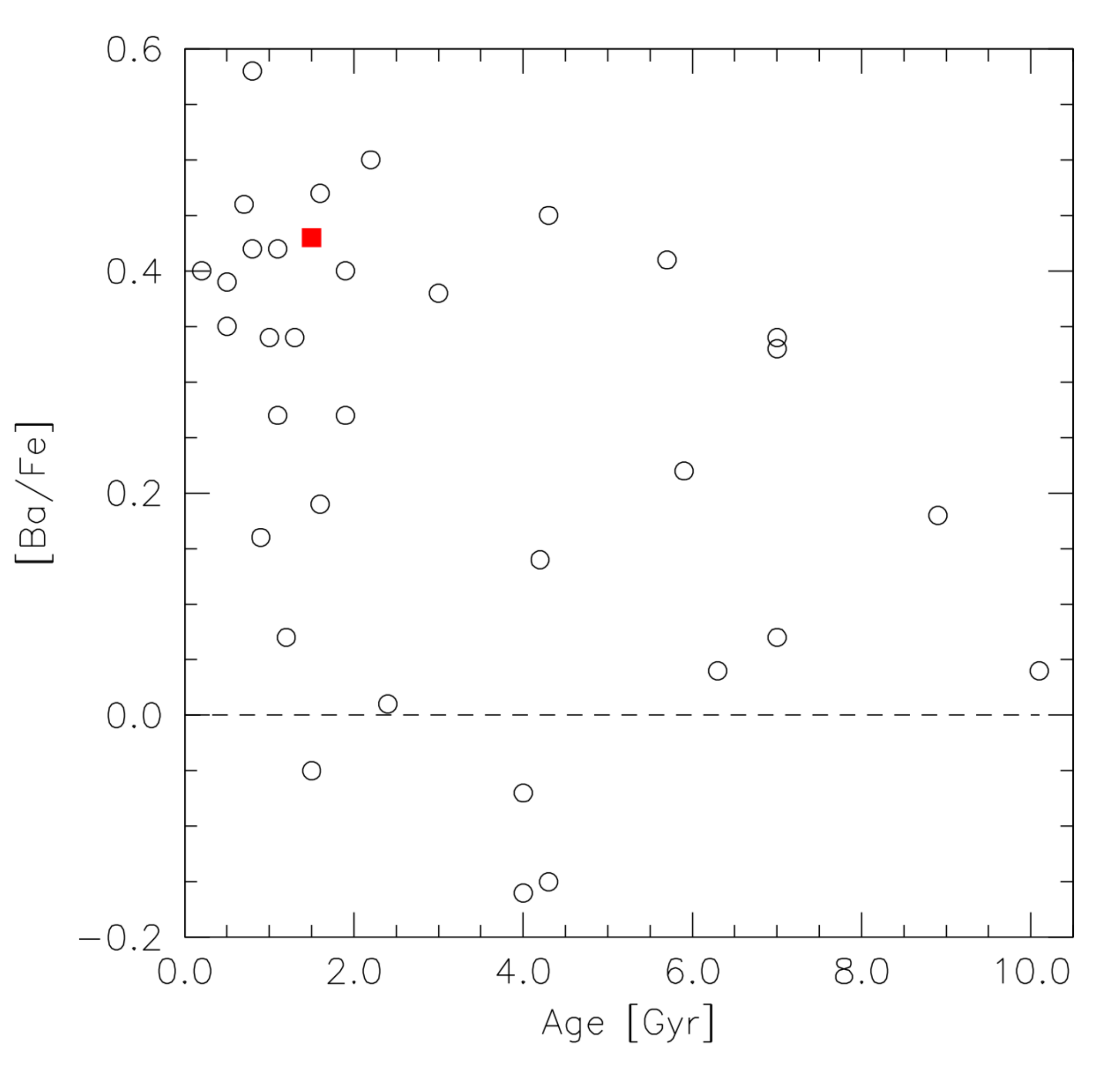}
\caption{Trend of [Ba/Fe]  abundance ratio versus metallicity (left panel) and versus age (right panel) for a compilation of old open clusters. The red square indicates NGC 7762}
\end{figure*}

\noindent
We have two $\alpha$ element abundance ratios (for Si and Ti)  in common with Curtis et al.(2013) and
Pakhomov et al. (2009).  Our mean [Ti/Fe] = 0.08 is compatible with their [Ti/Fe] =0.03 within the uncertainties, while our [Si/Fe]= 0.25 is clearly overabundant with respect to the Sun compare with their
[Si/Fe]=0.02. We remark however that the involved number of stars is relatively low and both studies might be affected by low number statistics. Overall the two clusters appear to share common properties. Only, NGC 772 is more distant from the Sun ($\sim$ 900 pc) than Ruprecht~147 ($\sim$ 300 pc).

\subsection{Iron peak elements}
Our wavelength coverage allows us to measure only one Iron-peak element: Ni. Our mean value for [Ni/Fe] is 0.14$\pm$0.08, while Curtis et al. (2013) found for Ruprecht~417 a value
of [Ni/Fe]=-0.02$\pm$0.05.  Given the low number statistics, we conclude that also this element is close to solar.

\subsection{Neutron-capture elements}
Mishenina et al. (2015, and references therein) showed that  Galactic old open  clusters for which Ba  was measured 
exhibit  a clear Ba overabundance with respect to the Sun. The overabundance correlates with age, since  it increases at decreasing age. 
In addition, for clusters younger than 3-4 Gyr, the Ba abundance shows a large scatter. In Fig.~5 we compare NGC 7762  [Ba/Fe] ratio with a compilation of old open clusters from Mishenina et al. (2013, 2015) and Jacobson \& Friel (2013). Although incomplete, this compilation has the advantage that the typical age and metallicity ranges for Galactic old open clusters are well covered. The comparison for NGC 7762 is shown in the two panels of Fig.~5, where our cluster is identifies with a red square. The left panel shows the well-know large scatter of [Ba/Fe] at any metallicity, and NGC 7762 comfortably sits among the other old open clusters. In the right panel, we show that the addition of NGC 7772 does not change the general scenario. The mean [Ba/Fe] overabundance from we measured from our 8 giant stars ([Ba/Fe]= 0.43$\pm$0.15) follows the general trend.\\

\section{Conclusion}
In this work we have presented the very first abundance analysis for the Galactic old open cluster NGC 7762. This cluster was is in fact poorly known, mostly because of its large extent and the severe
field star contamination. We observed 9 stars in the evolved region of the CMD, and isolate 8 members: 3 clump stars, two RGB stars and 3 possible AGB stars. 
Our results can be summarised as follows:\\

\begin{itemize}
\item Our spectroscopic analysis reveals that the cluster is of solar metal content ($[Fe/H]$=0.04$\pm$0.12), which we would expect from his location close to the Sun. 
\item Our best fit age is 2.5$\pm$0.5 Gyr,
which is mostly in line with previous - purely photometric - estimates.
\item With respect to previous investigations we argue in this study that the distance is slight larger, and position the cluster at 900 pc from the Sun. We confirm previous reddening estimates within the observational uncertainties.\\
\item Beside metallicity, we found that the abundance ratios we could measure also reveal a solar-scaled mixture of elements in the cluster. 
\item In NGC 7762 barium is over abundant with respect to the Sun. This result follows the trend of the general population of old open clusters in the Milky Way. \\
\end{itemize}

\noindent
Overall, the fundamental parameters we obtained in this study makes NGC 7762 a twin of Ruprecht 147, a coeval cluster located at only 300 pc from the Sun.

\acknowledgments
G. Carraro thanks the CAO staff, in particular Gennady Valyavin, for the kind hospitality, and G. Galazutdinov for his help during the preparation of this project.
The visit to CAO has been financially supported by the ESO Office for Science in Chile.
S. Villanova gratefully acknowledges support from the Chilean BASAL  and Centro de Excelencia en Astrofísica
y Tecnologías Afines (CATA) grant PFB-06/2007. We made extensive use of the WEBDA databased, maintained by E. Paunzen at Masaryk
University in Brno.

\vspace{5mm}
\facilities{CAO}

\software{MIDAS, IRAF}

\allauthors

\listofchanges

\end{document}